\documentclass[a4paper]{jpconf}
\usepackage{graphicx}
\usepackage{subfig}
\begin{document}
\title{Autoencoder-based Online Data Quality Monitoring for the CMS Electromagnetic Calorimeter}

 \author{Abhirami Harilal, Kyungmin Park, Michael Andrews\\ and Manfred Paulini on~behalf~of~the~CMS~Collaboration}
  \address{Department of Physics, Carnegie Mellon University, Pittsburgh, PA 15213, U.S.A.}

\ead{aharilal@cern.ch}

\begin{abstract}
The online Data Quality Monitoring system (DQM) of the CMS electromagnetic calorimeter (ECAL) is a crucial operational tool that allows ECAL experts to quickly identify, localize, and diagnose a broad range of detector issues that would otherwise hinder physics-quality data taking. Although the existing ECAL DQM system has been continuously updated to respond to new problems, it remains one step behind newer and unforeseen issues. Using unsupervised deep learning, a real-time autoencoder-based anomaly detection system is developed that is able to detect ECAL anomalies unseen in past data. After accounting for spatial variations in the response of the ECAL and the temporal evolution of anomalies, the new system is able to efficiently detect anomalies while maintaining an estimated false discovery rate between $10^{-2}$ to $10^{-4}$, beating existing benchmarks by about two orders of magnitude. The real-world performance of the system is validated using anomalies found in 2018 and 2022 LHC collision data. Additionally, first results from deploying the autoencoder-based system in the CMS online DQM workflow for the ECAL barrel during Run\,3 of the LHC are presented, showing its promising performance in detecting obscure issues that could have been missed in the existing DQM system.
\end{abstract}

\section{Introduction}

In a large-scale high energy physics experiment like CMS \cite{The_CMS_Collaboration_2008}, ensuring timely detection of issues that could affect the detector performance and the quality of data taken is a major task that requires significant personpower and time. Presently, the CMS online DQM~\cite{DQM2019_Azzolini}
consists of a set of histograms that are filled based on a first-pass analysis of a subset of data collected by the detector. These histograms are monitored continuously by a DQM \textit{shifter} who reports on any apparent irregularities observed. Depending on the severity of the problem, various mitigation measures up to stopping of the data taking would be performed by the relevant experts.
While this system has proven to be dependable, the changing running conditions and increasing collision rates, along with aging electronics, bring forth failure modes that are newer and harder to predict. Machine learning (ML) based approaches to anomaly detection in DQM have been adopted by previous efforts in CMS~\cite{ECAL_AE_2019, DT_AE_2019}. In this paper, an unsupervised method of anomaly detection for the online DQM of the CMS ECAL is presented utilizing an autoencoder (AE)~\cite{AE} on ECAL data processed as two-dimensional (2D) images.
In a novel approach, correction strategies are implemented to account for spatial differences in the ECAL response as well as the time-dependent nature of anomalies in the detector. This system is deployed in the online DQM for the ECAL barrel during LHC Run\,3 collisions, complementing the existing DQM plots. 
First results from the AE-based anomaly detection system are reported indicating it to be a highly valuable diagnostic tool for ECAL experts involved in real-time data taking operations. All the figures shown here are from the approved CMS public results in Refs.~\cite{CMS-DP-2022-043, CMS-DP-2023-002}.

\section{The CMS Electromagnetic Calorimeter and Data Quality Monitoring }

The CMS ECAL~\cite{ECAL} is a hermetic calorimeter which measures the energy, time, and position of photons, electrons and electromagnetic fraction of jets. It played a crucial role in the discovery of the Higgs Boson~\cite{higgs_discovery} as well as in the measurement of the Higgs properties. ECAL is made up of scintillating lead tungstate crystals arranged in a cylindrical central barrel (EB) section closed by two endcaps (EE+ and EE--).

\begin{figure}[tbh]
\centering{
\subfloat[]{\includegraphics[height=2.8cm,
width=0.38\textwidth]{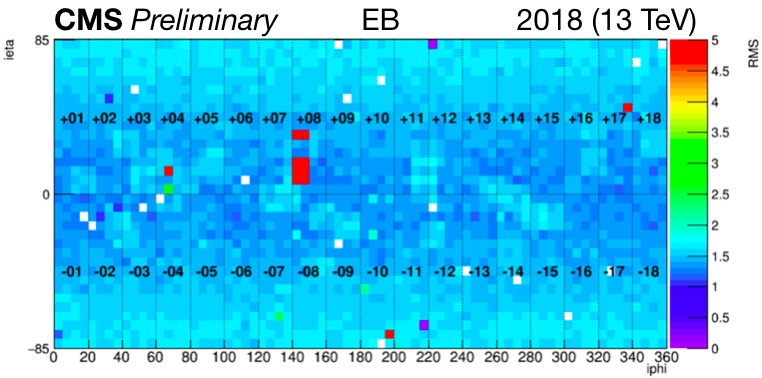}}
\hspace{0.5cm}
\subfloat[]{\includegraphics[height=2.8cm,
width=0.38\textwidth]{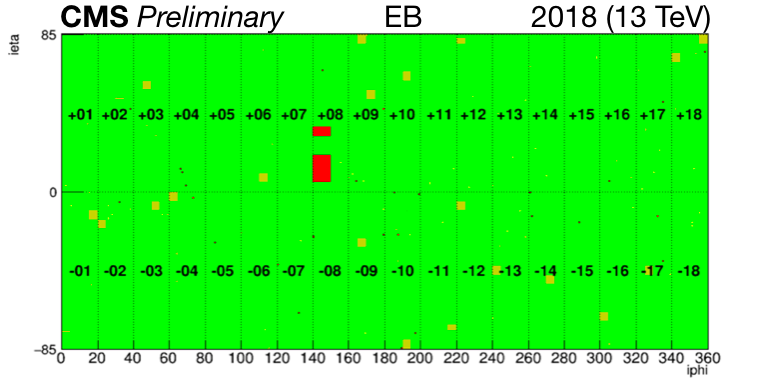}}
}
\caption{DQM plots for ECAL barrel: (a) occupancy-style and (b) quality-style histogram.}
\label{Fig:task}
\end{figure}

The DQM system in ECAL mainly consists of two kinds of histograms: occupancy-style histograms as shown in Fig.~\ref{Fig:task}(a), filled with real-time data of vital quantities, and quality-style histograms as shown in Fig.~\ref{Fig:task}(b), which are drawn based on thresholds and rules applied on the quantity in the occupancy-style histograms. Quality-style histograms are in color-coded maps that are easy-to-interpret, so that it is possible to tell at a glance if something is wrong in the ECAL. The color code used is as follows: green for \textit{good}, red for \textit{bad}, brown for a \textit{known problem}, and yellow for \textit{no data}, which may or may not be an issue depending on the context. The histograms are often plotted at the granularity of a \textit{trigger tower} (TT), which is defined as a set of $5\mathrm{x}5$ crystals. Each single square in Fig.~\ref{Fig:task} represents a TT, and 68 such TTs form a \textit{supermodule}~(SM) in the barrel, corresponding to the numbered rectangles in Fig.~\ref{Fig:task}.
In online DQM, these histograms are accumulated over a CMS data acquisition \textit{run}, with each histogram plotted at every \textit{lumisection}~(LS), that corresponds to a time interval of about 23 seconds, over which the luminosity is considered to remain approximately constant.

With increasing luminosity and harsher radiation environment, it becomes impossible to anticipate all failure modes in a complex detector like ECAL. Though there are multiple alarms set in the current DQM framework to catch various types of errors, they are prone to high false positives and are limited by having to define hard-coded rules for every possible detector geometry. To mitigate these issues, an automated approach to anomaly detection is explored using unsupervised ML.

\section{Autoencoder-based Anomaly Detection System}

\subsection{Network and Anomaly Detection Strategy}

An unsupervised anomaly detection method is developed using an AE trained on occupancy-style histograms from the ECAL DQM processed as 2D images. The AE built with a ResNet~\cite{ResNet} 
architecture using \textsc{Pytorch}~\cite{Pytorch-NEURIPS2019_9015} is trained with manually certified \textit{good} data. The encoder network of the AE encodes the input image into a lower dimensional latent space, and the decoder network tries to reconstruct the original image from the encoded space. The goodness of reconstruction is measured by the reconstruction loss $\mathcal{L}$, computed as the Squared Error between the input ($x$) and the AE-reconstructed output ($x'$) as defined in Eq.~\ref{eq:loss}:
\begin{equation}
\mathcal{L}(x,x') = |(x-x')|^{2}
\label{eq:loss}
\end{equation}
The network trained on good data reconstructs the nominal detector image well with minimal reconstruction loss. When fed with anomalous data, however, it fails to reconstruct the anomalies and gives a higher loss. The squared error on each tower is calculated and plotted as a 2D loss map, on which some post-processing steps are applied as explained in the next sections. A threshold is then derived from the anomalous loss values that can efficiently catch 99\% of the anomalies.

\subsection{Dataset, Training and Validation}
Each input image to the AE is the digitized hit occupancy map from a single LS. The dataset used for training and validation is taken from the 2018 runs during LHC Run\,2. It consists of 100\,000 2D occupancy images, with training and validation dataset split in 9-to-1 ratio. 
In order to make the quality interpretation consistent across different run conditions, occupancy maps are normalized with respect to \textit{pileup} (PU), which are additional proton-proton interactions within the same proton bunch crossing. After pre-processing with the PU correction, AE models are trained and validated separately for the barrel and each endcap.

In addition to the nominal validation, fake anomaly validation is performed, using the good images from the nominal validation set but with synthetic anomalies introduced in random towers. Three types of anomaly scenarios are studied:
\begin{enumerate}
    \item Missing SM/sector: Entire SMs for the barrel and sectors for endcaps are randomly set to have zero occupancy values in each LS.
    \item Single zero occupancy tower:  A single tower is set to have zero occupancy at random in each LS.
    \item Single hot tower: A single tower is set to be \textit{hot}, or having higher-than-nominal occupancy.
\end{enumerate}

\subsection{Spatial Response Correction}

\begin{figure}[tbh]
\centering{
\hspace{0.2cm}
\subfloat[]{\includegraphics[height=2.5cm,
width=0.32\textwidth]{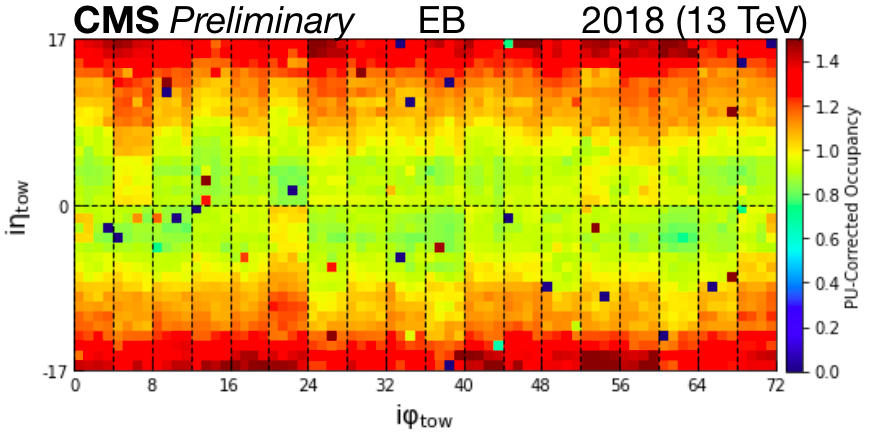}}
\subfloat[]{\includegraphics[height=2.5cm,
width=0.32\textwidth]{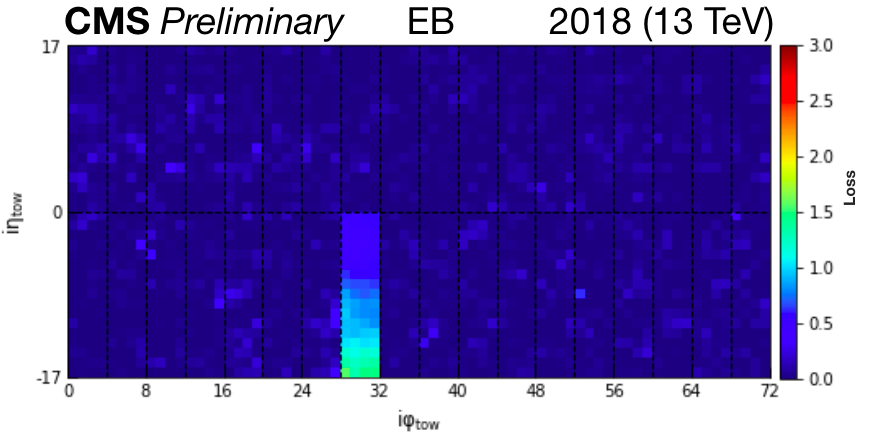}}
\hspace{0.2cm}
\subfloat[]{\includegraphics[height=2.5cm,
width=0.32\textwidth]{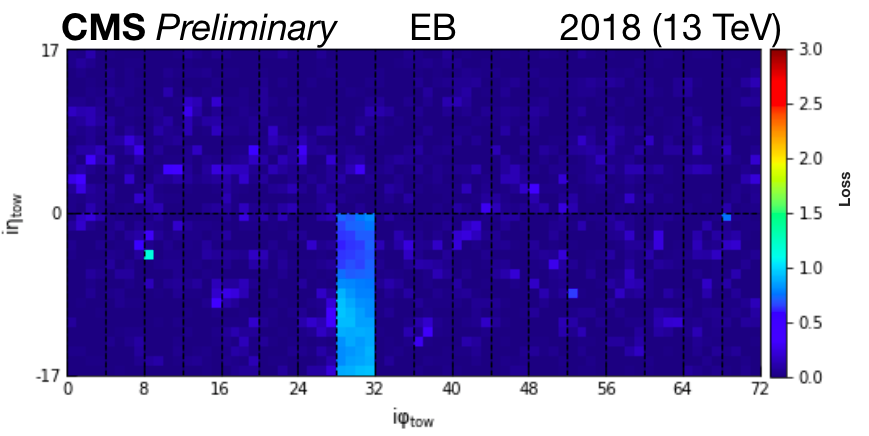}}
}
\caption{(a) Average occupancy map of EB. Loss maps for missing SM scenario (b) before and (c) after spatial response correction.}
\label{Fig:spatial}
\end{figure}

In the ECAL, the occupancy close to the beam pipe tends to be higher. This is clearly visible in the average occupancy map shown in Fig.~\ref{Fig:spatial}(a). This effect is reflected in the loss map shown in Fig.~\ref{Fig:spatial}(b) in the case of a missing SM, where the towers in the higher $|\eta|$ region have higher loss than those in the lower $|\eta|$ region, for an anomaly affecting all the towers in the SM equally. In order to interpret the towers as equally anomalous across the SM, the loss map is normalized by the average occupancy map, as a spatial response correction, to obtain a uniform loss map as illustrated in Fig.~\ref{Fig:spatial}(c).

\subsection{Time correction}

\begin{figure}[tbh]
\centering{
\includegraphics[width=0.8\textwidth]{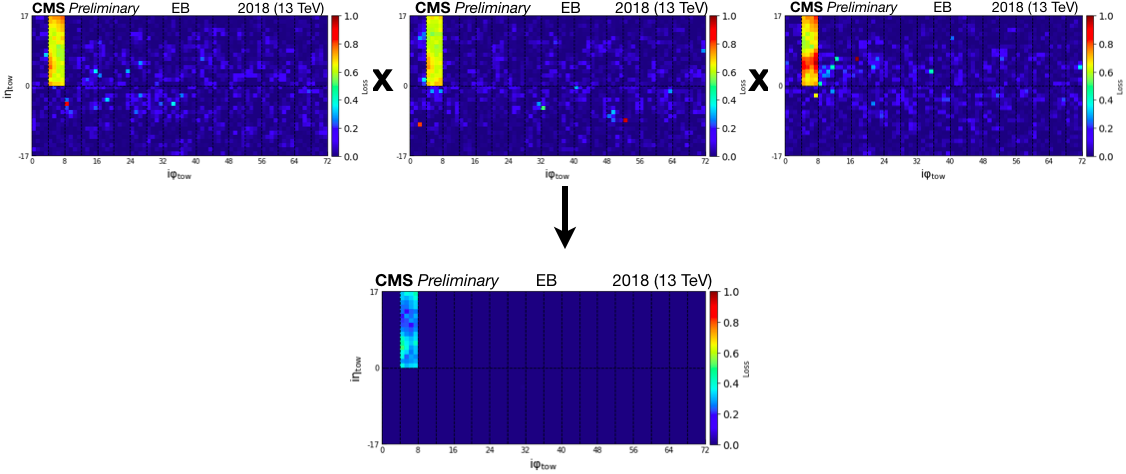} }
\caption{Time correction strategy: loss maps from three consecutive LSs (top panel) are multiplied at a tower level to make a time-corrected loss map (bottom).}
\label{Fig:timeCorr}
\end{figure}

A correction to further reduce false positives is implemented by utilizing the time-dependent nature of the anomalies in the detector. Unlike random fluctuations, real anomalies would persist with time. Figure~\ref{Fig:timeCorr} describes the time correction strategy, where loss maps that have been corrected for spatial effects from three consecutive LSs are multiplied with one another at a tower level to boost the anomaly and suppress the fluctuations.

\section{Results}
An anomaly tagging threshold is chosen based on the validation set with fake anomalies, such that 99\% of the anomalies are detected.
The performance of the AE is measured using the metric False Discovery Rate~(FDR) defined as:
\begin{equation}
    \textrm{FDR} = \frac{\textrm{Number of good towers above the anomaly threshold}}{\textrm{Number of good and bad towers above the anomaly threshold}}
\label{eq:FDR}
\end{equation}
The lower the FDR, lesser the false detection and better the performance of the AE.

\subsection{Testing on Fake Anomalies}

Table~\ref{tab:FDR} summarizes the FDR values from each anomaly scenario. For the barrel and the endcaps, single zero occupancy towers are most challenging to detect. Spatial correction improves the AE performance, with a much greater effect for the endcaps. This is due to the presence of a larger effective gradient in the occupancy values across the endcap region compared to the barrel. Further improvement in the FDRs by an order of magnitude is achieved with the time correction. Final performance scores after applying all post-processing are around 5\% for the zero occupancy tower scenario and sub-percent for the missing SM (for the barrel) and hot tower scenarios. Remaining false positives have likely come from real anomalous towers that have fallen into the manually certified good dataset.
Based on the performance for each scenario, a single anomaly tagging threshold is chosen for each sub-detector part, such that it can efficiently tag all anomalies.

\begin{table}[htb]
\centering{
\caption{Summary of FDR at 99\% anomaly detection in fake anomaly validation.}
\resizebox{0.95\textwidth}{!}{\begin{tabular}{|c|c|c|c|c|c|c|c|}
\hline
Scenario
&  \multicolumn{1}{c|}{Missing SM} 
&  \multicolumn{3}{c|}{Zero Occupancy Tower} 
&  \multicolumn{3}{c|}{Hot Tower}\\ \hline

 & Barrel & Barrel & EE+ & EE-- & Barrel & EE+ & EE-- \\ \hline
\begin{tabular}[c]{@{}c@{}}AE \\ no correction \end{tabular} & 3.6\% & 51\% & 86\% & 87\% & 2.8\% & 0.01\% & 0.00\% \\ \hline
\begin{tabular}[c]{@{}c@{}}AE after\\ spatial correction \end{tabular} & 3.1\% & 49\% & 13\% & 14\% & 2.9\% & 0.06\% & 0.05\% \\ \hline
\begin{tabular}[c]{@{}c@{}}AE after\\ spatial and time correction \end{tabular} & 0.13\% & 4.1\% & 5.6\% & 6.3\% & 0.00\% & 0.00\% & 0.00\% \\ \hline


\end{tabular}
\label{tab:FDR}
   }
 }
\end{table}

\subsection{Testing on Real Anomalies}

\begin{figure}[tbh]
\centering{
\subfloat[Input occupancy map of EB with an error in SM~EB$-$03]{\includegraphics[height=2.3cm,
width=0.3\textwidth]{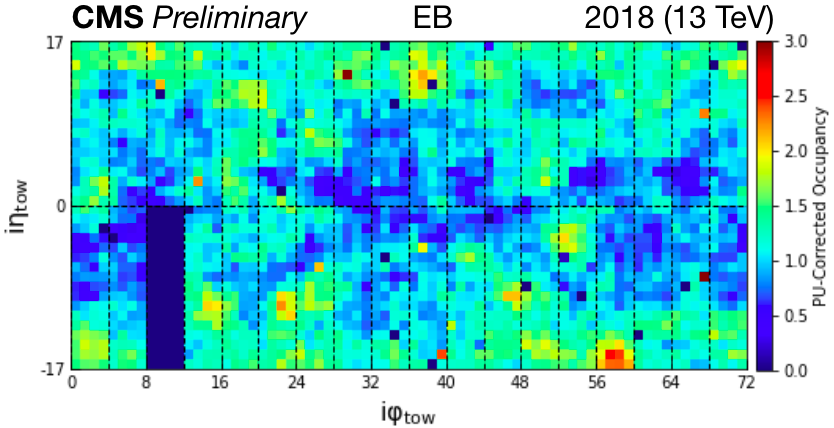}}
\hspace{0.2cm}
\subfloat[Input occupancy map of EB with a group of hot towers]{\includegraphics[height=2.3cm,
width=0.3\textwidth]{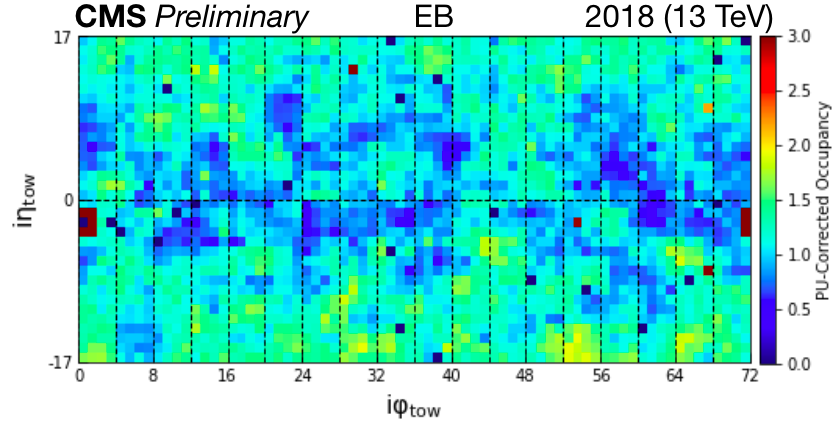}}
\hspace{0.3cm}
\subfloat[Input occupancy map of  EE with two zero occupancy towers]{\includegraphics[height=2.3cm,
width=0.17\textwidth]{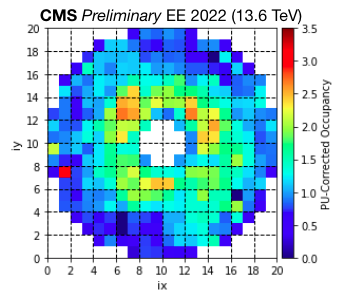}}
\\
\subfloat[Final AE quality plot with the bad SM in red]{\includegraphics[height=2.3cm,
width=0.305\textwidth]{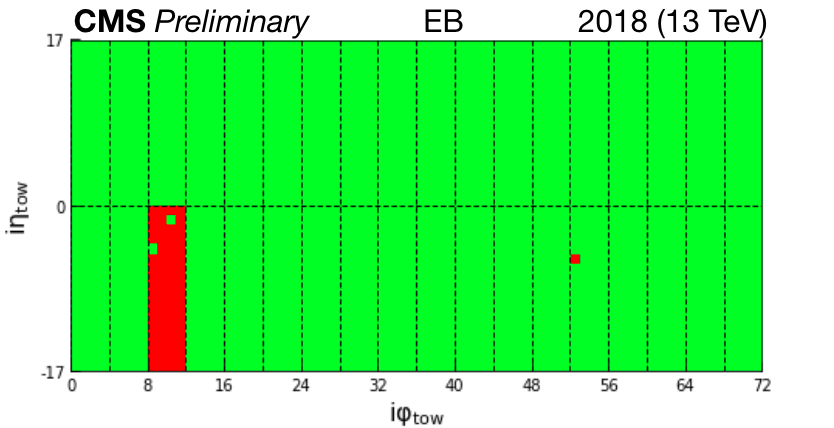}}
\hspace{0.2cm}
\subfloat[Final AE quality plot with the bad towers in red]{\includegraphics[height=2.3cm,
width=0.305\textwidth]{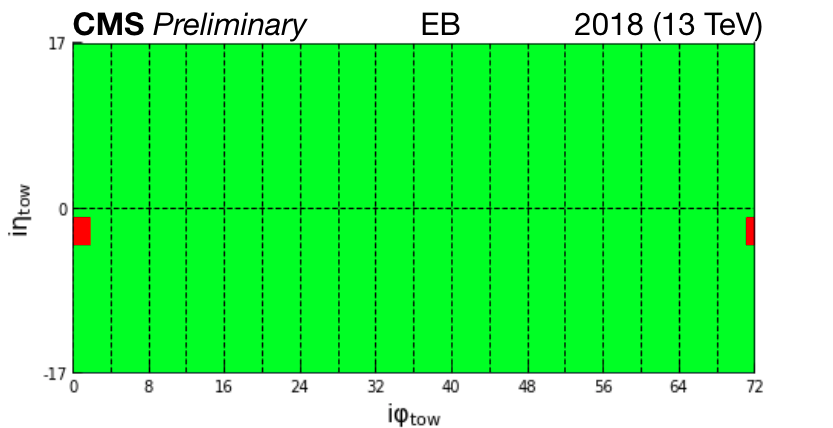}}
\hspace{0.25cm}
\subfloat[Final AE quality plot with the two bad towers in red]{\includegraphics[height=2.3cm,
width=0.16\textwidth]{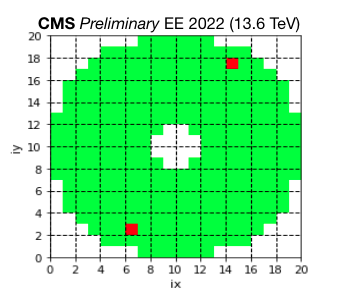}}
}
\caption{Input occupancy images with real anomalies (top) and corresponding AE quality plots (bottom) for the barrel and endcaps.}
\label{Fig:realanom}
\end{figure}

The AE model is further tested using real anomalous data from 2018 and 2022 LHC runs as illustrated in Fig.~\ref{Fig:realanom}.
Figures~\ref{Fig:realanom}(a) and (b) respectively show occupancy plots from Run\,2 with a missing SM~EB$-$03, and with a ring of hot towers with a zero occupancy tower in the center in~EB$-$01 and~EB$-$18.
Their corresponding final quality plots from the AE loss map identify the bad towers in red in Figures~\ref{Fig:realanom}(d) and ~\ref{Fig:realanom}(e) respectively.
For the endcaps, a real anomaly case from Run\,3 is illustrated in Fig.~\ref{Fig:realanom}(c), with two zero occupancy towers and a hot tower around the edge. The AE quality plot in Fig.~\ref{Fig:realanom}(f) spots the two zero occupancy towers correctly. The hot tower does not show up in the quality plot, as it is previously identified as problematic and therefore masked in the DQM system.

The performance of the AE on real-life anomalies demonstrates its ability to detect various kinds of anomalies using a single threshold, without the need for hard-coded rules based on the type of anomaly or the geometry, and emphasizes the power of unsupervised ML as an efficient, adaptable anomaly detection tool.

\section{Deployment in Online DQM for LHC Run3}
\begin{figure}[tbh]
\centering{
\subfloat[]{\includegraphics[height=2.1cm,
width=0.32\textwidth]{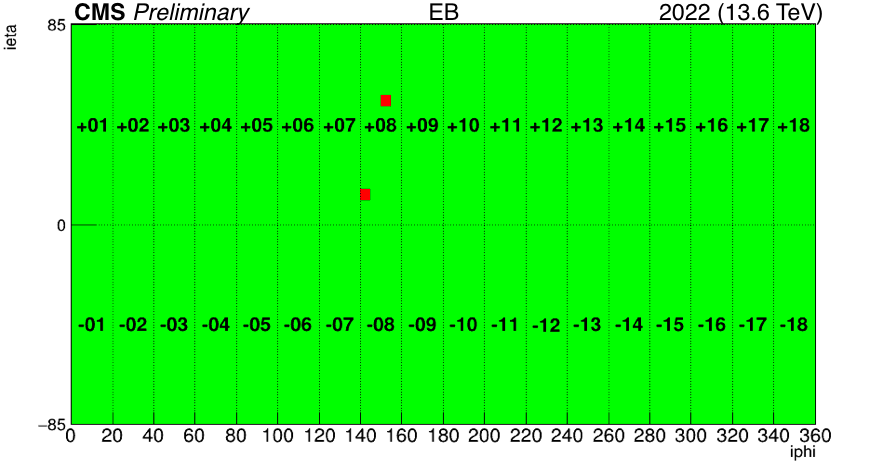}}
\hspace{0.1cm}
\subfloat[]{\includegraphics[height=2.1cm,
width=0.32\textwidth]{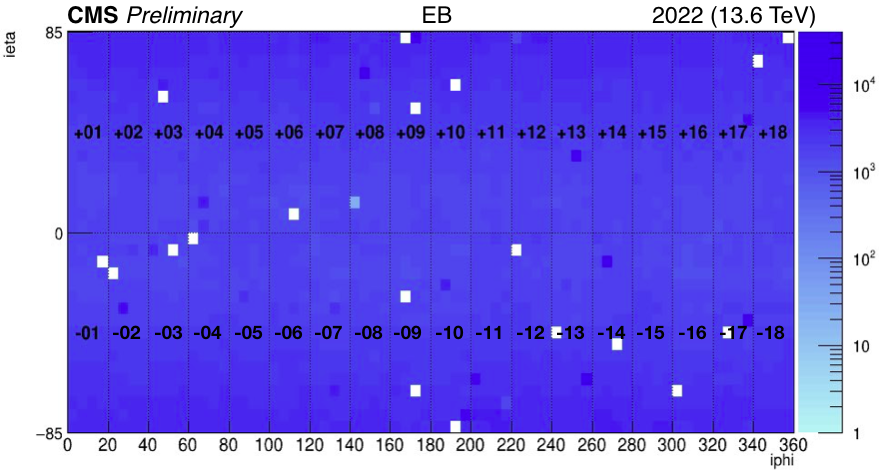}}
\hspace{0.1cm}
\subfloat[]{\includegraphics[height=2.5cm,
width=0.32\textwidth]{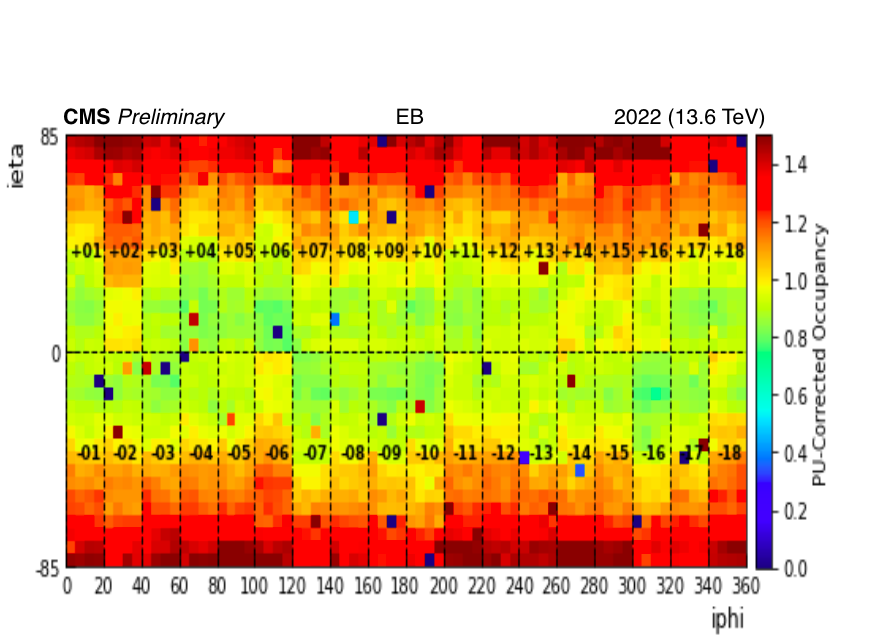}}
}
\caption{(a) ML quality plot accumulated over 9 LSs, from the MLDQM deployed in the online DQM workflow for EB from a 2022 run, showing two bad towers in red. (b) The digitized hit occupancy plot accumulated over all the LSs in the full run from the online DQM for the same run. (c) PU-corrected average occupancy plot over several runs in 2022.}
\label{Fig:deploy}
\end{figure}

The AE-based anomaly detection system, named MLDQM, has been deployed in the ECAL online DQM workflow for the ECAL barrel since the start of Run 3 of the LHC. This introduces a new ML quality plot from the AE in the ECAL DQM workspace (see Fig.~\ref{Fig:deploy}(a)) complementing the existing plots. The model inference is done by exporting the trained 
\textsc{Pytorch} model to \textsc{Onnx}~\cite{onnx} and run in CMS production using \textsc{Onnx Runtime}~\cite{onnxruntime}.
The deployment of the models for the endcaps is underway and will be ready for the 2023 runs.

It has been observed that the MLDQM is able to correctly identify consistently bad towers, as well as transient bad towers, which may point to deteriorating channels in the ECAL that are not easily detected by the existing DQM plots.
Figure~\ref{Fig:deploy}(a) shows the ML quality plot with two towers in~EB+08 marked red: Tower\,1 closest to $i\eta = 0$ and Tower\,2 closest to $i\eta = 85$. Figure~\ref{Fig:deploy}(b) shows the total digitized hit occupancy map accumulated over all LSs of the run in the online DQM. Here, Tower\,1 is visible with very low occupancy compared to other towers, indicating that it is a persistent zero occupancy tower, while Tower\,2 is faintly visible, indicating that it likely had zero occupancy in several LSs but not in all, thus pointing to a transient anomaly.
This feature also shows up in the occupancy map in Fig.~\ref{Fig:deploy}(c), produced offline by averaging over several runs in Run\,3. The low occupancy of Tower\,2 in this plot implies that the tower indeed had zero occupancy transiently for many LSs in these runs, and that it is likely not a stand-alone random occurrence. 
This tower could be pointing to channels on the way to becoming permanently faulty. This feature of the MLDQM which detects potentially deteriorating channels would be immensely helpful to detector experts monitoring the health of the detector. By keeping track of how often a particular tower is flagged as bad by the AE and defining a threshold on this frequency, experts can choose to mask the transient tower proactively.

\section{Summary}
A robust, deployment-ready AE-based anomaly detection and localization system has been developed for the CMS ECAL using unsupervised ML. An efficient pileup-based normalization strategy is applied on images made from the raw detector data, making quality interpretations consistent across changing experimental conditions. The application of novel techniques of spatial and time corrections
gives an order of magnitude improvement in performance. The autoencoder-based system demonstrates efficient detection of anomalies of various degrees, shapes, and positions in the detector up to a tower-level granularity, using a single efficient threshold for each sub-detector part. The MLDQM system deployed in the online DQM workflow for the ECAL barrel performs well on real-time data from Run\,3 by detecting anomalies as well as identifying potential degrading channels that goes undetected by the existing DQM plots. This ML-based DQM system is designed to complement and improve the existing DQM by helping detector experts make more accurate decisions and reduce false alarms. This autoencoder-based anomaly detection system can be generalized and adapted to other particle physics experiments for data quality monitoring.


\section*{References}
\bibliography{BibTeX/iopart-num}

\end{document}